# Serving Database Information Using a Flexible Server in a Three Tier Architecture

Herbert Greenlee,  Robert Illingworth,  Jim Kowalkowski, Anil Kumar,  Lee Lueking, Taka Yasuda, Margherita Vittone, Stephen White
*FNAL, Batavia, IL 60510, USA*

The DØ experiment at Fermilab relies on a central Oracle database for storing all detector calibration information. Access to this data is needed by hundreds of physics applications distributed worldwide. In order to meet the demands of these applications from scarce resources, we have created a distributed system that isolates the user applications from the database facilities. This system, known as the Database Application Network (DAN) operates as the middle tier in a three tier architecture. A DAN server employs a hierarchical caching scheme and database connection management facility that limits access to the database resource. The modular design allows for caching strategies and database access components to be determined by runtime configuration. To solve scalability problems, a proxy database component allows for DAN servers to be arranged in a hierarchy. Also included is an event based monitoring system that is currently being used to collect statistics for performance analysis and problem diagnosis. DAN servers are currently implemented as a Python multithreaded program using CORBA for network communications and interface specification. The requirement details, design, and implementation of DAN are discussed along with operational experience and future plans.

## 1. INTRODUCTION

The DØ experiment employs a three tier architecture to access information from a centralized database. This arrangement has many desirable features, isolating the client applications form the database server and schema. The Database Application Network (DAN) is the result of a project to build such a middle tier server, a second generation system, based on experience gained from operating DØ servers built for the Calibration and SAM[1,2] projects. It is designed to supporting thousands of repository accesses per hour for constants, configuration, and dataset information, and deliver data to all the client jobs in a timely fashion

### 1.1. Requirements

There are many requirements for the DAN server from the performance and scalability to its configuration. Using the middle tier removes any dependencies of applications to a particular database vendor database API.  It should provide low latency for multiple simultaneous client requests. It provides connection management and controlled access to the database. The application clients need little knowledge of the database schema and all queries are centralized. It is possible to establish remote proxy servers which maintain persistent caches and remain operational even during network and central database outages. The system should scale in a controlled fashion, and be easily tuned, configured, and administered. It is important to have detailed activity monitoring and error handling. The system must be able to respond to and provide information for all calibration data requests.

## 2. WHAT IS DAN?

The DAN system is a Python based server operating between the database and the user applications. This server performs database transactions on behalf of the user, and provides an application-level protocol for accessing detector calibrations. The hierarchical overview of the system is depicted in Figure 1 with the central database server shown at the top, and DAN servers deployed to provide the needed functionality and performance for many applications.

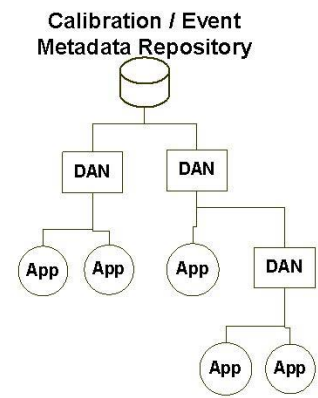

Figure 1: The hierarchical deployment of DAN servers which enables the performance and functionality required for all applications to access information from the central databases .

### 2.1. Features

The system has several important features which enable it to meet the requirements stated above. The layered architecture of the DAN server itself is shown in Figure 2. A user API provides connection to the client through a CORBA[3] framework.  It uses a multi-level cache strategy which offers fast delivery of commonly used objects and reduces the number of transactions to the central database server. The type of information is Write Once Read Many (WORM) and this caching avoids coherency problems.

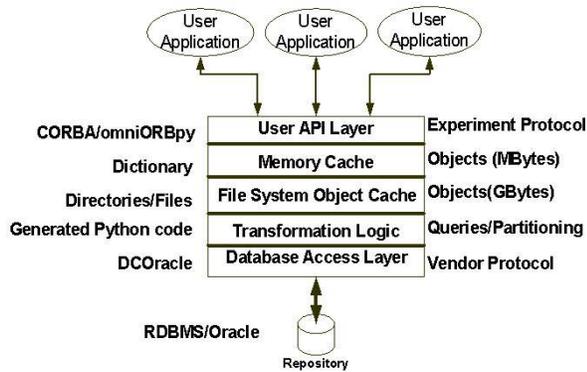

Figure 2: The DAN server architecture components.

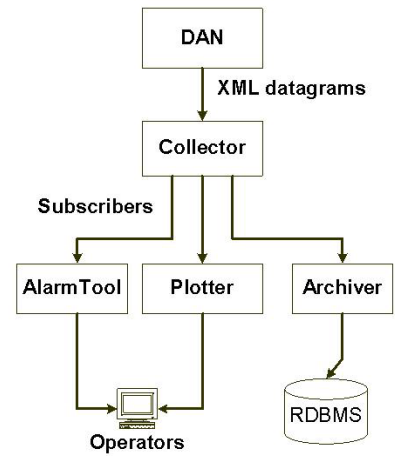

Figure 3: The DAN Monitoring configuration.

Database connection pooling allows many clients to use the same connection to the database, thus significantly reducing the number of concurrent users the database encounters. Requests for the same data are consolidated, greatly reducing the transaction rate. The system provides a code generator that uses the database schema to build the backend which accesses the database, and the C++ header files for the clients. The middle tier approach reduces maintenance and allows us to capture common usage patterns based on the statistics it provides.

Significant reduction in development time is afforded by built in table-to-object transformation policies through which calibrations are delivered in logical objects. Load balancing is possible by selectively assigning servers to heavy usage applications, and it is possible to manage the number of connections each server has to the central database and thus throttle the transactions when required. A proxy mode of operation is supported under which the middle tier servers can be chained together to provide additional scalability and reliability for the system. The server is multi-threaded further reducing latency and allowing simultaneous object caching and database querying.

## 2.2. Monitoring

To operate the system efficiently, and tune the configuration, it is essential to collect extensive monitoring statistics, and diagnostic information. Monitoring resource usage and activity is performed through a subscription service as illustrated in Figure 3. Verbose recording of interesting events (errors, informational, or debugging) is accomplished through a compact XML format. Event thresholds are based on asynchronous notifications providing a powerful problem solving aid.

## 3. TESTING AND OPERATION

Simple testing was performed to confirm the basic operation of the system, but it has been difficult to schedule adequate resources for doing good load testing. The calibration information accessed with the various servers ranges from a few words per query to get magnet polarities, to 800,000 rows of pedestals and gains for the Silicon Micro Tracker (SMT) device. In testing, requests for the SMT information were made from 50 client nodes. The size of the information requested is about 25 MB, although there is a large overhead for storing this in the Python cache as CORBA objects requiring almost 300MB of machine memory. The initial access from the database requires about 6.5 minutes to complete, while accessing the same information once it is cached is almost 4 times faster. While these tests were underway less than 30% of the CPU was being used for the DAN server on a dual Athlon 1800 Linux server. The network delivery rate was approximately 3.2MB/s.

### 3.1. Deployment

The initial deployment of the system is shown in Figure 4. The calibration servers have been initially deployed on two Linux nodes with a third reserved for failover as shown in Figure 4. Each node has dual Pentium 868 MHz processors and 1 GB of RAM and 2 GB of swap space. The operating system is Fermi Linux which is based on Red Hat Linux Kernal 2.4.9-31. Each production node is designated for a specific user activity including analysis and reconstruction.

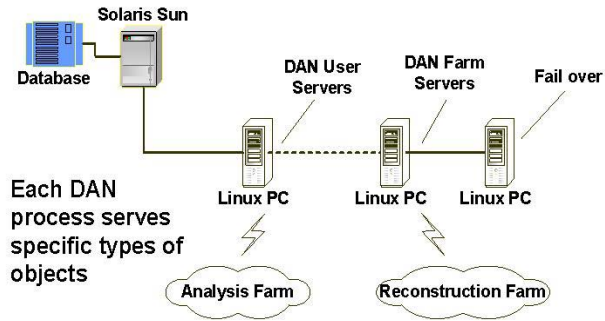

**Figure 4:** The DØ DAN Deployment strategy includes Linux servers for the reconstruction farm, and user clients. A failover server box provides the required 24/7 level of service.

In the event of a node failure, the servers will be auto-started on the failover node with notification sent to the supporting email list.  This enables the system to be employed for 24/7 operation and has functioned effectively. There are certain modes of failure which have not triggered the correct failover behavior and the system is still being perfected.

Typical transaction rates vary from 300 to 3000 requests per hour,  with transaction size ranging from 1.8kb to 4.5kb, dependent on the table accessed in the data base, for the server tested. We anticipate that the number of active, high demand, servers operated at FNAL will be around 20, with approximately 6 to 10 additional sites beyond Fermilab which run high demand DAN proxy servers with disk cache.

## 4. EVALUATION AND PLANS

### 4.1. CORBA, Python, and Other Packages

There are several areas where product or design choices were made including the use of CORBA, Python, and several third party software packages. CORBA was chosen for historical reasons, it being widely used for distributed object brokering in the DØ data management system. However, it is not an optimal choice for calibration data access which is not a very object oriented task, and CORBA data representations are not very useful in this application. Python is an excellent choice for rapid development and scripting. It greatly simplifies third party product integration, and because it is platform independent, it provides ease of deployment.

Python, however, has several drawbacks that make it less suitable for a production environment. Python has very high memory requirements, and slower execution speed than some other language choices. Although it offers multi-threading this feature is limited to one CPU, which is a severe restriction when using a dual processor machine.  The debugging tools for Python, in particular for multi-threaded applications, are rather poor.

The DAN server employs several third party software packages which have added to the ease of programming, but have also introduced some problems. DC Oracle[4] provides a Python API to Oracle has had poor backward compatibility between versions introducing additional debugging and support effort. OmniORBpy[5] is a python CORBA implementation providing many nice features, and a significant improvement over Fnorb previously used, introduced random thread lockup problems which were very difficult to locate.

One feature of the DAN product is code generation to create the base classes needed to describe the table structure in the database. This introduced a single edit point for code changes to multiple classes, and is difficult to understand  and maintain.

### 4.2. Future Plans

The existing server is now being used with a C++ cache management tandem component which  reduces the memory for one SMT run set from 300MB down to about 25MB. This significantly improves the performance and allows 20 or more SMT run sets to be cached simultaneously in memory without severely taxing the server machine, or slowing the delivery to multiple simultaneous clients. Distributing the code and configuring the servers is complicated because of the mixed-language mode of operation. Furthermore, operation of the tandem servers is more complex than the single python server. Nevertheless, this seems to be an adequate solution to provide the experiments needs for the next few months.

Re-writing the server completely in C++ is a  viable option which maintains the current client interface while providing all the performance improvements of the tandem solution, plus this solution provides additional robustness and performance in the secondary disk cache and database connection management. It is felt that most of the design in the current Python server could be used and a C++ product would map almost class per class to the existing Python code. Instead of DCOracle, now used in the Python code as the database backend, we would need to use an OCI or ODBC interface. What statistics are needed in a C++ server and if the monitoring points be the same are still open questions.

Another interesting alternative is to employ web services to provide the needed functionality. DØ has been using a system of flat files to distribute their SMT calibration information before the SMT server was deployed. This is somewhat awkward because all of the calibration files need to be distributed to the reconstruction processing machines. However, it has the distinct advantage that because the data is well understood by the experiment, it can be carefully packed into the files, reducing the size to only about 1 MB per SMT set. It would be very convenient to have a straightforward system that would construct and deliver this information on demand from the database. We have discussed this option and feel it would be quite straightforward to build such a system to explore these ideas.

In the coming months we will complete the statistics gathering, remote monitoring, and controls package, as well as provide additional graphs via the web. Features for changing resource allocations or other configuration parameters dynamically are being included which will provide the ability to more easily tune the system in real time. More extensive load testing will be performed to further evaluate the performance of the system, and we will observe the system in production on the reconstruction farms and for user analysis.

## Acknowledgments

Many people have contributed significantly to this project. We thank Vicky White who initiated the project with a set of carefully drafted requirements and significant input to the design and planning. We thank those at DØ, on the reconstruction farms and general users, who helped evaluate the system and have accommodated many problems while this project was being debugged. We owe special thanks to Harry Melanson, Michael Diesburg, and Slava Kulik who have worked with us to test and commission the product. This work is sponsored by DOE contract No. DE-AC02-76CH03000.